\documentclass[aps]{revtex4}
\input{amssym}

\def\t {\tilde}
\def\be {\begin{equation}}
\def\ee  {\end{equation}}
\def\bea {\begin{eqnarray}}
\def\eea {\end{eqnarray}}
\def\nn {\nonumber}

\begin{document}

\fontsize{12}{18} \selectfont

\title{On singularity resolution in quantum gravity}

\author{Viqar Husain$^*$ and Oliver Winkler$^\dagger$}

\affiliation{$^*$Department of Mathematics and Statistics,\\
University of New Brunswick, Fredericton, NB E3B 5A3, Canada.  \\
$^\dagger$Perimeter Institute of Theoretical Physics\\
Waterloo, ON Canada.\\
EMail: husain@math.unb.ca, owinkler@perimeterinstitute.ca\\
\mbox{ }}

\date{Dec. 20, 2003}

\begin{abstract}
\fontsize{12}{16} \selectfont We examine the singularity
resolution issue in quantum gravity by studying a new quantization
of standard Friedmann-Robertson-Walker geometrodynamics. The
quantization procedure is inspired by the loop quantum gravity
programme, and is based on an alternative to the Schr\"odinger
representation normally used in metric variable quantum cosmology.
We show that in this representation for quantum geometrodynamics
there exists a densely defined inverse scale factor operator, and
that the Hamiltonian constraint acts as a difference operator on
the basis states. We find that the cosmological singularity is
avoided in the quantum dynamics. We discuss these results with a
view to identifying the criteria that constitute "singularity
resolution" in quantum gravity.
\end{abstract}



\maketitle


\section{Introduction}

It is widely believed that a quantum theory of gravity will give
insights into the question of what becomes of classical curvature
singularities. This is based largely on intuition from uncertainty
principle and fundamental length scale arguments in regions of
large spacetime curvatures. What is required to address the
problem quantitatively is quantization of model systems that
contain classical metrics with curvature singularities. Such
models are usually symmetry reductions of general relativity or
other generally covariant metric theories. Within a model an
obvious approach is to look at classical observables such as
curvature scalars, and see if they can be represented as operators
on a suitable Hilbert space. Their spectra and quantum dynamics
may give an indication of what becomes of the classical
singularity.

This question has been studied using models derived from symmetry
reductions of general relativity since the late 1960's
\cite{misner1,bb1,bb2,vh1}. All of this work used the
Arnowitt-Deser-Misner (ADM) (metric variable) Hamiltonian
formulation of general relativity ("geometrodynamics") as the
classical starting point, and the Schr\"odinger representation as
the quantum starting point for developing a quantum gravity model.
The results obtained from various mini- and midi-superspace models
were largely inconclusive. Some indicated singularity avoidance,
others did not, but no insights emerged as general and definitive
in the sense of transcending the model studied.

After the development of the Ashtekar (connection variable)
Hamiltonian formulation ("connection dynamics"), many of the
questions studied in the ADM formulation were revisited, including
the general canonical quantum gravity program (for reviews see
\cite{carlo,thomas}). The different classical variables led to the
development of a non-Schr\"odinger representation program based on
holonomy variables ("loop quantum gravity"). Recently results from
this programme were applied by Bojowald \cite{mb1,mb2} to the old
question of quantizing mini-superspace models, with a view to
studying what happens to classical curvature singularities upon
quantization. This application has produced some interesting
results: for Friedman-Robertson-Walker (FRW) mini-superspace
models the Hamiltonian constraint acts like a difference operator
on the space of states, and there is an upper bound on the
spectrum of the inverse scale factor operator. Taken together,
these results lead to the conclusion that the big bang singularity
is resolved in the loop quantum gravity approach \cite{abl}.

A number of questions may be asked at this stage concerning
classical variables, quantization procedures, and singularity
resolution: What criteria constitute singularity avoidance? Is the
singularity avoidance conclusion from the loop quantum gravity
programme a result of both the classical starting point {\it and}
the choice of representation? Would a non-Schr\"odinger
representation quantization in the geometrodynamical ADM variables
give the same results?

Motivated by these questions, we study a new quantization of flat
FRW cosmology (the model in which the loop quantum gravity results
were first obtained \cite{mb1,mb2}). Our classical starting point
is the geometrodynamical Hamiltonian formulation, which we
quantize via a non-Schr\"odinger representation motivated by
holonomy-like variables. We obtain results qualitatively similar
to those obtained in loop quantum cosmology: the Hamiltonian
constraint acts like a difference operator on the space of states,
and the spectrum of the inverse scale factor has an upper bound.

This paper is organised as follows: The next section introduces
the classical theory and a set of basic variables for FRW
geometrodynamics. Section III describes the quantization procedure
and discusses the volume, inverse scale factor, and Hamiltonian
constraint operators. The final section contains a discussion of
the results with the singularity resolution question in mind.

\section{Classical theory}

The canonical Hamiltonian variables in geometrodynamics are
$(q_{ab}, \tilde{\pi}^{ab})$, where the configuration variable
$q_{ab}$ is the metric on a spatial 3-surface, and the conjugate
momentum variable $\t{\pi}^{ab}$ is a function of the surface's
extrinsic curvature. In terms of these variables, the vacuum
Hamiltonian and spatial diffeomorphism constraints are
\bea
 H &=& {1\over \sqrt{q}}\left(\t{\pi}^{ab}\t{\pi}_{ab}
      - {1\over 2}\t{\pi}^2\right) -\sqrt{q}R = 0 \label{hcons}\\
{\cal C}_a &=& D_b\t{\pi}^b_a = 0,
\eea
where $q=$det$q_{ab}$, $\t{\pi}= \t{\pi}^{ab}q_{ab}$, and $R$ is
the Ricci scalar of $q_{ab}$.

For the flat FRW model a reduced Hamiltonian theory may be
obtained by the ansatz
\be
  q_{ab} = B(t) e_{ab}, \ \ \ \ \ \ \ \  \t{\pi}^{ab} = P(t) e^{ab}
\ee
where $e_{ab}$ is the flat Euclidean metric diag(1,1,1). The
density weight on $\t{\pi}^{ab}$ is obtained using this fiducial
metric. Unlike in general relativity where $B(t)$ is dimensionless
since it is a metric variable, we will take it to have dimension
length square (which means that the spatial coordinates are
dimensionless). The conjugate momentum $P(t)$ is then
dimensionless in order that the symplectic form
\be
  \omega = {1\over 8\pi G_N}\ dB \wedge dP
\ee
has dimensions of action (in $c=1$ units). The phase space
topology is that of a particle on the half line, $\Bbb R^+\times
\Bbb R$, since $B(t)>0$ in this parametrization. The standard FRW
scale factor is $a(t) = \sqrt{B(t)}.$

The Hamiltonian constraint is
\be
H = -{3\over 2} P^2 \sqrt{B}
\ee
and the diffeomorphism constraint ${\cal C}_a$ vanishes
identically. It is interesting to note that this Hamiltonian
constraint is identical in form to that obtained in the connection
variables \cite{mb1}.

To facilitate quantization however, it is useful to make a
canonical transformation that lifts the half line restriction. We
proceed in two steps. The first is to extend the classical
configuration space to include the singularity $B(t)=0$. This step
is essential for addressing the question of singularity resolution
for without it there is "no classical singularity to avoid" in the
quantum theory. (This for example is done -- more or less
unconsciously --  in the familiar quantization of the hydrogen
atom by the choice of Hilbert space $L^2(\Bbb R_0^+)$, even though
the classical configuration space is only $\Bbb R^+$.) The second
step is to reparametrize the configuration variable such that its
domain is the real line.

This is done by introducing new variables $(x,p)$ by  writing
$x^2=B$. The symplectic form becomes
\be
\omega = {1\over 8\pi G_N}\ 2x\ dx\wedge dP = {1\over 8\pi
G_N}\ dx \wedge d(2xP).
\ee
Thus the new momentum is $p=2xP$, for which $\{x,p\}=8\pi G_N.$
(Both coordinates $x$ and $p$ now have dimension length.) Note
that in this parametrization, the point $x=0$ is included to give
the full real line as the configuration space. This amounts to an
extension of the original parametrization to include the
degenerate metric with $B(t)=0$. (This feature is also present in
the connection-triad variables, where invertibility of the triad
is not a requirement.)

The Hamiltonian constraint (\ref{hcons}) as a function of $(x,p)$
is
\be
 H = -{3\over 8}\ {p^2\over x^2}\ |x| = -{3\over 8}\ {p^2\over
|x|}.
\ee
Note that this constraint is now quite different in form than the
one in the connection-triad variables, although the $x$ variable
may be regarded as a "triad." The reason for this is that the new
momentum $p$ is a product of the metric and extrinsic curvature
variables, and this is quite unlike the connection variable $A$
made of the triad connection $\Gamma$ and extrinsic curvature $K$
as $A=\Gamma + K$. (It is this connection $A$ that is essential
for formulating the loop representation through holonomies
\cite{carlo,thomas}.)

We now introduce an algebra of classical observables and write
quantities of physical interest as functions of these variables.
Their form is motivated by the holonomy observables used for
quantization in the loop quantum gravity programme. We use as the
fundamental variables $x$ and
\be
U_\gamma(p) := {\rm exp}(i\gamma p/L),
\ee
where $\gamma$ is a real parameter and $L$ fixes the (arbitrary)
unit of length. The parameter $\gamma$ is necessary in order to
separate momentum points in phase space (eg. fixing $\gamma/L =1$
say, gives the same value of $U$ for $p$ and $p+2n\pi$). This
variable may be viewed as a {\it momentum analog of the holonomy
variable} of loop quantum gravity.

The pair $(x, U_\gamma(p))$ has the Poisson bracket algebra
\be
\{x, U_\gamma(p)\} =  (8\pi G_N)\ {i\gamma \over L}\
U_\gamma(p).
\label{pb}
 \ee
This is the basic algebra which will be taken over to the quantum
theory.

Other quantities of interest are the volume, which up to a
multiplicative constant is
\be
V(t) = B(t)^{3/2} = |x|^3,
\ee
and inverse powers of the scale factor $a(t)$. The latter has
proven useful in determining whether there is "singularity
avoidance" in a quantum theory \cite{mb1}. It is possible to write
these observables, and the Hamiltonian constraint using our choice
of fundamental variables $(x, U_\gamma(p))$.

The standard FRW scale factor is
\be
a(t) = \sqrt{B(t)} = |x|.
\ee
Using the method introduced in \cite{tt2}, inverse powers of the
scale factor may be defined classically either via the bracket
\be
U_\gamma^{-1} \{ V^n,  U_\gamma \} = U_\gamma^{-1} \{
|x|^{3n}, U_\gamma \}
 = i (8\pi G_N)\ {\gamma\over L}\ 3n\ {\rm sgn}(x) |x|^{3n-1},
\label{inva}
\ee
or more simply by inverse powers of the volume observable. However
the latter definition cannot be carried over to the quantum theory
since, as we see will below, the volume operator turns out to have
zero as an eigenvalue, so negative powers of it are not densely
defined. Therefore the somewhat indirect definition (\ref{inva}),
with $n>0$ will be useful in studying the approach to the
singularity in the quantum theory. The requirement that the power
of $x$ on the right hand side be negative means that $n<1/3$. Thus
we need $0 < n < 1/3$ in order to obtain a sequence of inverse
powers of the scale factor in terms of the basic variables. The
choice $n=1/6$ gives
\be
 {{\rm sgn}(x)\over \sqrt{|x|}} = -{2Li\over (8\pi G_N)\gamma}\ U_\gamma^{-1}
 \{ V^{1/6}, U_\gamma \},
\label{scale-1}
\ee
and $n=1/4$ gives
\be
{1\over |x|} = \left({4\over 3(8\pi G_N)\gamma\t{L}}\right)^4\
\left( U_\gamma^{-1} \{V^{1/4}, U_\gamma \} \right)^4
\label{scale-2}
\ee
From this it is clear that powers of Poisson bracket in Eqn.
(\ref{inva}) may be used as a starting point for defining a large
class of operators for inverse powers of the scale factor. It is
of interest to see whether these all lead to qualitatively similar
behavior concerning the quantum nature of the big bang
singularity.

The Hamiltonian constraint can be written as a function of the
basic variables by using the relation (\ref{scale-1}) for the
inverse scale factor as follows:
\be
 H = -{3\over 8}\ {p^2\over |x|} = {3L^2\over 2(8\pi G_N)^2\gamma^2}\ p^2
\ \left(U_\gamma^{-1} \{ V^{1/6}, U_\gamma \} \right)^2
 \ee
Note here that there is the alternative choice of using
(\ref{scale-2}) to define the Hamiltonian constraint. However,
this leads to a more complicated form due to the larger number of
$U$ factors.

It is of course also possible to write the Hamiltonian constraint
in other classically equivalent ways. One alternative is to
substitute $|x|/x^2$ rather than directly using the inverse scale
factor $1/|x|$ from (\ref{scale-1}) or (\ref{scale-2}). These
choices will clearly lead to inequivalent operators in the quantum
theory since the number of factors of $U$ are different. As for
the inverse scale factor, it is generally useful to also study
various choices for the Hamiltonian constraint in order to
identify the main features that are to be associated with
"singularity resolution" in the quantum theory.

In the following, we focus primarily on the simplest ordering of
the Hamiltonian constraint, and the square of the definition
(\ref{scale-1}) for the inverse scale factor, since both of these
contain the smallest number of $U$ factors. However we will
briefly comment on other choices.

\section{Quantum theory}

To construct the quantum theory for the classical system described
above, we will proceed in analogy to the procedure used in loop
quantum gravity. The first step is to choose an algebra of
classical functions that is represented as quantum configuration
operators. We take here the algebra generated by the functions
\be
W(\lambda) = e^{i \lambda x/L},
\ee
where $\lambda \in \Bbb R$. It consists of all functions of the
form
\be
f(x) = \sum_{j=1}^n  c_j\ e^{i \lambda_j x/L},
\ee
with $c_j \in \Bbb C$ and their limits with respect to the
supremum norm. This algebra is known as the algebra of almost
periodic functions over $\Bbb R$ and we denote it by $AP(\Bbb R
)$.

It is well-known that $AP(\Bbb R)$ is naturally isomorphic to
$C(\overline{\Bbb R}_{Bohr})$, the algebra of continuous functions
on the so-called Bohr-compactification of $\Bbb R $
\cite{bratteli-robinson}. As the name suggests, $\overline{\Bbb
R}_{Bohr}$ is a compact group which can be obtained as the dual
group of ${\Bbb R}_{discr}$, the real line endowed with the
discrete topology. This suggests that taking
$L^2(\overline{\Bbb{R}}_{Bohr},d\mu_0)$, where $\mu_0$ is Haar
measure on $\overline{\Bbb R}_{Bohr}$, as the Hilbert space for
our theory is a viable option. {\it This is the decisive point
where we depart from the traditional approach in geometrodynamics,
where the Hilbert space is the conventional Schr\"odinger space}
$L^2(\Bbb R,dx)$. Once we adopt this new choice, basis states in
our Hilbert space are given by
\be
|\lambda \rangle \equiv | e^{i \lambda x/L} \rangle, \ \lambda
\in \Bbb{R},
\ee
with the inner product
\be
\langle \mu \mid\lambda \rangle = \delta_{\mu,\lambda}.
\ee
This representation has been discussed in some mathematical detail
in \cite{halvorson}, and also in \cite{polymer} where it is
applied to the quantization of a particle. Notice the difference
from the standard quantum mechanics of a particle on the real
line, where the right hand side is given instead by a delta
function $\delta(\mu - \lambda)$. This feature is traceable to the
fact that the configuration space is the real line with the
discrete topology, which in turn stems from the choice of the
algebra of functions.

The action of the configuration operators $\hat{W}(\lambda)$ is
defined by
\be
\hat{W}(\lambda)|\mu \rangle = e^{i\lambda \hat{x}/L}|\mu
\rangle = e^{i \lambda \mu}|\mu \rangle.
\ee
It is straightforward to verify that these operators are weakly
continuous in $\lambda$, which procures the existence of a
self-adjoint operator $\hat{x}$, acting on basis states according
to
\be
\hat{x} |\mu \rangle =  L\mu |\mu \rangle. \label{xop}
\ee

The next step is to construct the operators corresponding to the
classical momentum functions $U_\gamma = e^{i \gamma p/L}$. Their
action on the basis states is fixed by the definition of the
$\hat{x}$ operator and the requirement that the commutator between
$\hat{x}$ and $\hat{U}_\gamma$ reflects the corresponding Poisson
bracket (\ref{pb}) between $x$ and $U_\gamma$. With the definition
\be
\hat{U}_\gamma |\mu \rangle = |\mu -\gamma \rangle,
\ee
the commutator is
\be [\hat{x},\hat{U}_\gamma] = - \gamma L \hat{U}_\gamma
\ee
Making now the standard commutator-Poisson bracket correspondence
$[,] \leftarrow\rightarrow i\hbar \{,\}$, gives using (\ref{pb})
the relation
\be
-\gamma L = i\hbar \ (8\pi G_N) {i\gamma\over L},
 \ee
which fixes the length $L$ to $L=\sqrt{8\pi}l_P$. This shows
explicitly how the eigenvalues of $\hat{x}$ arise in Planck units
upon quantization.

Obviously, $\hat{U}_\gamma$ is unitary, however, it is not weakly
continuous with respect to $\gamma$. As a consequence, {\it there
is no momentum operator in this representation}, in stark contrast
to the Schr\"odinger quantization.

With the  basic quantum operators now at our disposal, we are in a
position to construct the inverse scale factor operator and
investigate its spectrum.

\subsection{\large Volume and Inverse Scale Factor}

The operator for the volume $\hat{V}$ is provided directly by the
operator $\hat{x}$ defined in (\ref{xop}). We have
\be
\hat{V} |\mu \rangle = (\sqrt{8\pi} l_P)^3|\mu|^3 |\mu
\rangle.
\ee
The operators corresponding to $U$ and $V$ can be used to obtain
an operator for the inverse scale factor. One way to do to this is
to use the square of the expression in Eqn. (\ref{scale-1}) with
$\gamma=1$. The resulting operator is
\be
\label{invop}
\widehat{{1\over |x|}} := {1\over 2\pi l_P^2}
\left( \hat{U}^{-1} [\hat{V}^{\frac{1}{6}}, \hat{U} ] \right)^2.
\ee
The key question is whether this operator is unbounded as in
standard quantum cosmology, where its eigenvalues diverge when
approaching the quantum state corresponding to $a=0$, or whether
it is bounded, indicating a (kinematical) resolution of the
classical singularity. To decide this we calculate its eigenvalue
on a basis state $|\mu \rangle$:
\bea
\label{invopeigen}
\widehat{ {1\over |x|}}\ |\mu \rangle
&=&
{1\over 2\pi l_P^2}\
\left[\hat{U}^{-1}\left(\hat{V}^{\frac{1}{6}}\hat{U}
- \hat{U} \hat{V}^{\frac{1}{6}}\right)\right]^2 |\mu \rangle\nn\\
& = & {1\over 2\pi l_P^2}\ \left( \hat{U}^{-1}
\hat{V}^{\frac{1}{3}}\hat{U}
                   -\hat{U}^{-1} \hat{V}^{\frac{1}{6}}
\hat{U} \hat{V}^{\frac{1}{6}} - \hat{V}^{\frac{1}{6}} \hat{U}^{-1}
\hat{V}^{\frac{1}{6}}
\hat{U} + \hat{V}^{\frac{1}{3}} \right) |\mu \rangle \nn\\
& = & {\sqrt{2\over \pi l_P^2}}\ \left( |\mu -1| - 2
|\mu|^{\frac{1}{2}}|\mu
-1|^{\frac{1}{2}} +|\mu| \right) |\mu \rangle \nn\\
& = & {\sqrt{2 \over \pi l_P^2}} \left(|\mu|^{\frac{1}{2}} - |\mu
-1|^{\frac{1}{2}}\right)^2 |\mu \rangle. \eea
This result reveals some important properties of the eigenvalues.
First, they are always positive or at most zero, as should be the
case. Second and more importantly, the spectrum is clearly bounded
from above. For $|\mu| \to \infty$ the eigenvalues approach $0$,
as would be expected from the behavior of $1/|x|$ for large $|x|$.
Moreover, the eigenvalue of the state $|\mu = 0\rangle$
corresponding to the classical singularity ($\hat{a}|0\rangle
\equiv \widehat{|x|}|0\rangle =0$) is $\sqrt{2/\pi l_P^2}$, and
this is the largest possible eigenvalue. (This is notably
different from the results in \cite{abl}, where the eigenvalue of
the inverse scale operator for the state $|\mu=0\rangle$ is $0$,
and the maximal eigenvalue is obtained instead for the state
$|\mu=1\rangle$. Although there are no principal reasons why this
could not happen in the quantum regime, it seems somewhat
unnatural from the classical point of view. It should be pointed
out however that this result is obtained in our formalism for a
different choice of operator ordering.)

In summary, this new quantization of the inverse scale factor in
geometrodynamics mimics the expected classical behavior for large
values $a(t)$, and departs significantly from the divergence in
the standard quantization near the classical singularity $a(t)=0$.
In this sense, the quantization resolves the singularity. This
"resolution" however is so far only kinematical, since we have not
investigated the quantum dynamics. It is conceivable that the
quantum dynamics breaks down at the state $|0 \rangle$, in which
case it would be hard to claim a satisfactory resolution of the
singularity. As the dynamics is encoded in the Hamiltonian
constraint, we now turn our attention to its operator realization.

\subsection{\large Hamiltonian constraint}

As discussed already in the classical section, the Hamiltonian can
be written in many different, classically equivalent forms. The
one we will focus on in this section is
\be
\label{H}
H = - \frac{3}{8}\frac{p^2}{|x|},
\ee
as this is in some sense the simplest one, and the spectrum of the
inverse scale operator is already known. As $p$ does not exist as
an operator in our quantum representation, we have to choose an
alternative way to represent $p^2$ as an operator. One way to do
this is motivated by the classical expression
\be
   p^2 = L^2\lim_{\gamma\to 0}\ {1\over \gamma^2}\ \left(2-U_\gamma -
   U_\gamma^{-1}\right).
\ee
A physical interpretation of this expression is obtained by
setting $\gamma = l_F/L_{phys}$ where $L_{phys}$ is the
characteristic size of the system under consideration, and $l_F$
is a fundamental length scale. (Note that a Hamiltonian naturally
introduces a scale $L_{phys}$ for a physical system.) The limit
then suggests that the "point" form of the momentum is recoverable
in the case $L_{phys}>>l_F$.

For quantum cosmology these considerations mean $l_F=l_P$ and
$\gamma = l_P/L_{phys}$, and lead to a Hamiltonian constraint
operator
\bea \label{Hop} \hat{H}_{\gamma} &=& \frac{3\pi l_P^2}{\gamma^2}\
\left(\hat{U}_{\gamma}
  + \hat{U}^{-1}_{\gamma} -2\right)    \widehat{ {1\over |x|}}  \nonumber\\
&=&  \frac{3}{2\gamma^2} \left(\hat{U}_{\gamma}
  + \hat{U}^{-1}_{\gamma} -2 \right)
 \left(\hat{U}^{-1} [ \hat{V}^{1/6} , \hat{U} ] \right)^2,
\eea
where a specific operator ordering has been chosen. The action of
$\hat{H}_{\gamma}$ on a basis state is given by
\bea
\label{hact}
\hat{H}_{\gamma} |\mu \rangle  &=&
\frac{\sqrt{18}}{\gamma^2}\ l_P \left(|\mu|^{1/2} -
|\mu-1|^{1/2}\right)^2
\left( |\mu +\gamma \rangle +|\mu - \gamma \rangle -2 |\mu \rangle \right)  \nn\\
&\equiv& \frac{\sqrt{18}}{\gamma^2}\ l_P\mathcal{V}(\mu)\
\left(
|\mu +\gamma \rangle +|\mu - \gamma \rangle -2 |\mu \rangle
\right)
\eea
On the eigenstate $|0\rangle$ of volume with zero eigenvalue,
which is the classical singularity, we have
\bea
\hat{H}_\gamma |0\rangle &=& \frac{\sqrt{18}}{\gamma^2}\ l_P
( |\gamma\rangle +|-\gamma\rangle - 2|0\rangle), \\
\widehat{ {1\over |x|}}\ |0 \rangle &=& \sqrt{2\over \pi l_P^2}\
|0\rangle
\eea
These equations represent the effects of quantization on the
classical singularity. In order to probe the dynamical part
further we must solve the quantum constraint equation that encodes
time evolution.

As is well known in the theory of constrained systems,
normalizable solutions of the quantum constraints do not lie in
the kinematical Hilbert space $\mathcal{H}$, but rather in a
larger space $\mathcal{C}^{\star}$. This space can be obtained as
the dual space of the dense subspace $\mathcal{C}$ of
$\mathcal{H}$, which is spanned by all elements of the form
\be
\sum_{i=1}^{n} \psi(\mu_i) |\mu_i \rangle.
\ee
A general element of $\mathcal{C}^{\star}$ can thus be written as
\be
\langle \psi | = \sum_{\mu} \psi(\mu) \langle \mu |.
\label{gend}
\ee
Notice that, while the sum is continuous as it runs over every
real number, its action on an element of $\mathcal{C}$ is well
defined by construction. The constraint equation -- symbolically
written as
\be
\hat{H}|\psi \rangle = 0,
\ee
is now interpreted as an equation in the dual space,
\be
\langle \psi | \hat{H}^{\dagger} =0.
\ee
Using the form of a general element of the dual space (\ref{gend})
and the (dualized) action of the (dual) Hamiltonian on (dual)
basis elements, we can derive a relation for the coefficients
$\psi(\mu)$:
\be
\label{psi}
\mathcal{V}(\mu + \gamma)\psi( \mu +\gamma ) - 2
\mathcal{V}(\mu) \psi (\mu ) + \mathcal{V}(\mu - \gamma) \psi (\mu
-\gamma) =0.
\ee

What is the meaning of this equation and in what sense does it
encode the quantum dynamics? First of all, it determines the
coefficients for those dual states that are physical. As in the
classical theory solutions to the constraint equation represent
classical spacetimes, these physical dual states can be
interpreted as representing "quantum spacetimes".

The difference equation (\ref{psi}) gives  physical states are
linear combinations of a countable number of components of the
form
\be
\psi (\mu + n\gamma ) |\mu + n\gamma \rangle,
\ee
where $\gamma $ is fixed at the Planck scale
($\gamma=l_P/L_{phys}\sim 1$) and $n\in \Bbb Z$. As each component
corresponds to a different eigenvalue for the volume and scale
factor, it can be interpreted as the quantum state representing
the universe at the "time" $\mu + \gamma$. A solution of the
Hamiltonian constraint therefore represents a linear combination
of FRW universes specified at certain discrete volumes, or
equivalently, at discrete times. It is in this sense that time
evolution is "discrete with fundamental time step" $\gamma$. It is
also clear that this "discrete evolution" does not represent the
state of a single universe at different discrete times, since the
term "single universe" has no meaning here. Rather a "discrete
solution" of the Hamiltonian constraint, (ie. one satisfying
(\ref{psi})), gives the amplitudes that the physical universe is
in one or other of the discretely separated components of the
physical state.

The state $<0|$ corresponding to the classical singularity is
contained in only one specific "quantum spacetime," (ie. solution
of the Hamiltonian constraint). Furthermore, in that one case we
can see that the system evolves right through the singularity
without encountering any problems, since the component $\psi(0)$
can be computed in terms of the components $\psi(\gamma)$ and
$\psi(-\gamma)$. In all other physical states, the state $<0|$
does not occur, and so in a sense one can say that the discrete
evolution "jumps" over the singularity if the state contains
components with both positive and negative values of $\mu$. In
such cases there is an instance of smallest but finite volume.

From these observations one can perhaps conclude that dynamically
the singularity has been resolved. A dynamical non-resolution of
the singularity might have occurred had it turned out that the
difference equations have no solutions if they contain the
$\psi(0)$ component, or if they contain components with both
positive and negative $\mu$ in the sum (\ref{gend}).

Finally, it is interesting to note that for our representation of
$\hat{H}$, the state at the classical singularity $\psi(0)$ can be
determined in contrast to the results in \cite{abl}. However, had
we chosen to write the classical Hamiltonian using eqn.
(\ref{scale-2}) instead of eqn. (\ref{scale-1}), which amounts to
using double the number of $U$ operators, we would have ended up
with the same result: $\psi(0)$ cannot be determined from the
difference equation, but a solution is still possible as it turns
out that $\psi(\gamma)$ is then given in terms of $\psi(-\gamma)$.
{\it This shows the significant differences that can arise due to
quantization ambiguities}. Ultimately, only physical predictions
and comparison with know facts or (as yet hypothetical)
experiments can determine the "right" choice.

\section{Conclusions and Discussion}

Our main result is that there is an alternative to the
Schr\"odinger quantization of the FRW cosmology in the standard
ADM geometrodynamical variables. This quantization leads to
conclusions qualitatively similar to those obtained in loop
quantum cosmology starting from the connection-triad variables:
(i) the Hamiltonian constraint acts like a difference operator,
and (ii) the inverse scale factor can be represented as a densely
defined operator. Thus {\it it is the representation space and the
realizations of the basic observables rather than the nature of
the classical variables} that are responsible for the similar
conclusions for this model.

To what extent is the quantization we have presented different
from the one employed in loop quantum cosmology? The differences
at the classical level are clear: the phase space variables
$(x,p)$ are not the standard mini-superspace variables that arise
via standard reduction from the connection-triad canonical
variables, as comparison with \cite{mb1,mb2} shows. The key
difference at the quantum level is that $\hat{U}_\gamma$ {\it is
not the holonomy operator} associated with the Ashtekar-Sen
connection for the FRW model. Rather, the $U_\gamma$ we use is a
standard translation generator whose realization on  the Hilbert
space $L^2(\overline{\Bbb{R}}_{Bohr},d\mu_0)$ is applicable to any
classical theory, as has been discussed in \cite{halvorson}. Thus
interpreting our quantization as a "loop representation" would
mean that one is generalizing this term to include {\it all
quantizations} on the Hilbert space
 $L^2(\overline{\Bbb{R}}_{Bohr},d\mu_0)$.

It is clear that the alternative representation based on the Bohr
compactification is applicable to other mini-superspace models,
since in all such models the phase space variables are functions
of only a time coordinate. It is also clear that this
applicability is independent of whether the classical phase space
variables are metric-extrinsic curvature or connection-triad. The
main difference between the variables arises in the form and
action of the Hamiltonian constraint.

In the flat FRW case we have discussed, the Ricci scalar term in
the Hamiltonian constraint vanishes. Thus the action of the
constraint as a difference operator is due only to the kinetic
term. In other mini-superspace models the Ricci scalar term, which
is a purely configuration variable, will have non-trivial action
on the basis states. However in the Bohr representation, this
action is multiplicative. Thus it appears that in other
mini-superspace models the "difference operator" feature of the
Hamiltonian constraint will survive. Similarly it appears that  an
inverse scale factor operator is definable using volume and $U$
operators, and that it is likely to have a spectrum bounded above.
For models where the phase space is more than two-dimensional, the
new representation can clearly be used for each pair of phase
space variables. Extension beyond mini-superspace (quantum
mechanics) to midi-superspace (quantum field theory) models, such
as the Gowdy cosmology would be of much interest \cite{ov2}.

"Singularity resolution" appears to consist of two main features,
one kinematical and the other dynamical. The kinematical feature
is the spectrum of the operator associated with a curvature scalar
(or other relevant classical observable) that diverges at a
curvature singularity. If the spectrum is bounded, the singularity
may be considered kinematically resolved. It is important to
identify the largest eigenvalue and corresponding eigenstate of
such an operator, since this is the "closest" the quantum theory
can get to the singularity. The dynamical feature of singularity
resolution concerns the action of the Hamiltonian constraint on
the state of largest curvature: this could lead either to no
solution of the constraint for zero or negative values of $\mu$,
or to a well defined "evolution" through  zero to negative values
of $\mu$. The former may be taken as an indication of the
breakdown of quantum evolution, and hence a dynamical
non-resolution of the singularity, regardless of the boundedness
of the curvature operator.

An alternative viewpoint is that the kinematical vs. dynamical
views are artificial in that the question of singularity
resolution is relevant only for the physical state space with a
well defined physical inner product. The question then becomes
whether there are any physical states for which the curvature
operator spectrum is unbounded. This appears more compelling, but
it has not been addressed here, or in the context of loop quantum
cosmology.

\bigskip
\noindent{\bf Acknowledgements}: We thank Thomas Thiemann for
discussions. V. H. thanks the the National Science and Engineering
Research Council of Canada for support.


\end{document}